\documentclass[conference]{IEEEtran}

\IEEEoverridecommandlockouts

\usepackage{cite}

\usepackage{amsmath,amssymb,amsfonts,mathrsfs,bm}

\usepackage{graphicx}
\usepackage{subfigure}
\graphicspath{{figures/}}

\usepackage{tabularx}
\usepackage{booktabs}

\makeatletter
  \newif\if@restonecol
\makeatother

\usepackage[linesnumbered,ruled,lined]{algorithm2e}
\usepackage{algpseudocode}

\usepackage{flushend}

\usepackage{scalerel}
\usepackage{tikz}
\usetikzlibrary{svg.path}

\definecolor{orcidlogocol}{HTML}{A6CE39}
\tikzset{
  orcidlogo/.pic={
    \fill[orcidlogocol] svg{M256,128c0,70.7-57.3,128-128,128C57.3,256,0,198.7,0,128C0,57.3,57.3,0,128,0C198.7,0,256,57.3,256,128z};
    \fill[white] svg{M86.3,186.2H70.9V79.1h15.4v48.4V186.2z}
                 svg{M108.9,79.1h41.6c39.6,0,57,28.3,57,53.6c0,27.5-21.5,53.6-56.8,53.6h-41.8V79.1z M124.3,172.4h24.5c34.9,0,42.9-26.5,42.9-39.7c0-21.5-13.7-39.7-43.7-39.7h-23.7V172.4z}
                 svg{M88.7,56.8c0,5.5-4.5,10.1-10.1,10.1c-5.6,0-10.1-4.6-10.1-10.1c0-5.6,4.5-10.1,10.1-10.1C84.2,46.7,88.7,51.3,88.7,56.8z};
  }
}

\newcommand\orcidicon[1]{\href{https://orcid.org/#1}{\mbox{\scalerel*{
\begin{tikzpicture}[yscale=-1,transform shape]
\pic{orcidlogo};
\end{tikzpicture}
}{|}}}}

\usepackage{fancyhdr}
\fancypagestyle{arxiv}{
    \fancyhead[L]{\vspace{0.2in}\footnotesize\textsf{Accepted by 2022 IEEE Intelligent Transportation Systems Conference (ITSC)}}
    
    \fancyfoot[C]{\thepage}
}

\fancypagestyle{arxivmain}{
    \fancyhead{}
    
    \fancyfoot[C]{\thepage}
}

\usepackage[colorlinks,allcolors=black]{hyperref} 

\title{\vspace{0.25in} \LARGE \bf Adaptive Testing for Connected and Automated Vehicles with\\ Sparse Control Variates in Overtaking Scenarios}

\author{Jingxuan Yang\textsuperscript{\orcidicon{0000-0001-9798-7347}}, Honglin He\textsuperscript{\orcidicon{0000-0003-4673-5283}}, Yi Zhang\textsuperscript{\orcidicon{0000-0001-5526-866X}}, \IEEEmembership{Member,~IEEE}, Shuo Feng\textsuperscript{\orcidicon{0000-0002-2117-4427}}, \IEEEmembership{Member,~IEEE}\\ and Henry X. Liu\textsuperscript{\orcidicon{0000-0002-3685-9920}}, \IEEEmembership{Member,~IEEE}%
\thanks{This work is supported by National Key Research and Development Program under Grant 2021YFB2501200 and National Natural Science Foundation of China under Grant 62133002. \textit{(Corresponding author: Shuo Feng.)}}%
\thanks{Jingxuan Yang and Honglin He are with the Department of Automation, Tsinghua University, Beijing 100084, China (email: \{yangjx20, hehl21\}@mails.tsinghua.edu.cn).}%
\thanks{Yi Zhang is with the Department of Automation, Beijing National Research Center for Information Science and Technology (BNRist), Tsinghua University, Beijing 100084, China (e-mail: zhyi@tsinghua.edu.cn).}%
\thanks{Shuo Feng is with the University of Michigan Transportation Research Institute, Ann Arbor, MI 48109, USA (e-mail: fshuo@umich.edu).}
\thanks{Henry X. Liu is with the Department of Civil and Environmental Engineering, University of Michigan, Ann Arbor, MI 48109, USA (e-mail: henryliu@umich.edu).}%
}

\begin{document}

\maketitle
\thispagestyle{arxiv}
\pagestyle{arxivmain}

\begin{abstract}

  Testing and evaluation is a critical step in the development and deployment of connected and automated vehicles (CAVs). Due to the black-box property and various types of CAVs, how to test and evaluate CAVs adaptively remains a major challenge. Many approaches have been proposed to adaptively generate testing scenarios during the testing process. However, most existing approaches cannot be applied to complex scenarios, where the variables needed to define such scenarios are high dimensional. Towards filling this gap, the adaptive testing with sparse control variates method is proposed in this paper. Instead of adaptively generating testing scenarios, our approach evaluates CAVs' performances by adaptively utilizing the testing results. Specifically, each testing result is adjusted using multiple linear regression techniques based on control variates. As the regression coefficients can be adaptively optimized for the CAV under test, using the adjusted results can reduce the estimation variance, compared with using the testing results directly. To overcome the high dimensionality challenge, sparse control variates are utilized only for the critical variables of testing scenarios. To validate the proposed method, the high-dimensional overtaking scenarios are investigated, and the results demonstrate that our approach can further accelerate the evaluation process by about 30 times.
\end{abstract}


\section{Introduction}

Testing and evaluation are major challenges for the development and deployment of connected and automated vehicles (CAVs). The past few years have witnessed increasingly rapid advances in the field of testing scenario library generation (TSLG) \cite{li2021scegene,wang2021advsim,menzel2018scenarios,tian2018deeptest,li2016intelligence,li2018artificial,li2019parallel,feng2020safety,feng2021parti,feng2021partii,zhao2016accelerated,zhao2017accelerated}. The goal of TSLG is usually to purposely generate safety-critical testing scenarios that can improve the evaluation efficiency of CAVs while ensuring the evaluation unbiasedness. As CAVs are usually black boxes, to evaluate the criticality values of different testing scenarios, surrogate models (SMs) are usually constructed by leveraging prior knowledge of CAVs. Due to the various types of CAVs, however, the performance dissimilarities between SMs and the CAVs under test usually exist, which may compromise the effectiveness of the testing scenarios and decrease the evaluation efficiency. Therefore, how to adaptively test different types of CAVs becomes a critical problem. 

Towards addressing this problem, several adaptive testing methods have been proposed \cite{mullins2018adaptive,koren2018adaptive,sun2021adaptive,feng2022adaptive}. The basic idea of these existing methods is to adaptively adjust the testing scenarios by leveraging the testing results of CAVs during the testing process. With more testing results of CAVs, more posteriori knowledge of CAVs can be obtained, and therefore the testing scenarios can be more customized and optimized for the CAVs under test, which can adaptively improve the testing efficiency \cite{feng2022adaptive}. However, most existing adaptive testing methods have limitations in dealing with complex scenarios, because of the ``Curse of Dimensionality'' (CoD) problem. For example, the CAV overtaking scenarios with two background vehicles could have 12-dimensional states (lateral and longitudinal positions and velocities of all three vehicles) and 3-dimensional actions (i.e., accelerations). Although the overtaking scenarios seem simple, if they last for 10 seconds at a frequency of 10 Hz, the dimension of the scenarios could exceed 10\textsuperscript{3}, which cannot be handled by most existing adaptive testing methods. 

The goal of this paper is to develop adaptive testing methods for high-dimensional scenarios using control variates (CVs). Instead of adaptively generating testing scenarios, we evaluate CAVs' performances by adaptively utilizing the testing results. The CVs are some random variables with means known \cite{owen2013monte}, which usually correlate the performance index of the event of interest. Through the regression of control parameters associated with the CVs, the testing results can be adaptively adjusted into a much narrower interval, which can greatly reduce the estimation variance. The worse the prior knowledge of CAVs is, the worse the testing results are, and the more they could be improved by the CVs. However, there also exists the CoD problem if we directly apply traditional CVs, since the number of control parameters would increase exponentially with the dimension of the scenarios.

To address the CoD problem, we propose the adaptive testing with sparse control variates (ATSCV) method in this paper to apply sparse control variates (SCVs) to the testing results, which only controls the variates associated with the critical variables of testing scenarios. To identify the critical variables, we applied the naturalistic and adversarial driving environment (NADE) \cite{feng2021intelligent}, which can generate high-dimensional testing scenarios by adjusting the critical variables of the naturalistic driving environment (NDE). Specifically, the critical variables are the scenes where NADE makes adversarial adjustments to the maneuvers of background vehicles, the dimension of which is much smaller than that of the scenarios. By generating control variates for the sparse critical variables, the SCVs can address the CoD problem for the adaptive testing method. Then, the SCVs are determined by multiple linear regression (MLR) techniques \cite{marill2004advanced,olive2017multiple}, which can essentially adjust the testing results. As the MLR techniques can leverage the posteriori knowledge of CAVs, the adjusted testing results can better evaluate the performance of the CAVs under test, resulting in an adaptive testing method. We note that this method is complementary to the adaptive testing methods that can adaptively generate testing scenarios during the testing process. To validate the proposed ATSCV method, the overtaking scenarios are investigated. Comparing with the estimation efficiency in NADE, the new adaptive testing method can further accelerate the evaluation process by about 30 times.

The rest of this paper is organized as follows. Section II introduces the overtaking scenarios and presents the testing scenario library generation methods in NDE and NADE. In Section III, the adaptive testing with sparse control variates method is proposed. Section IV evaluates the accuracy and efficiency of the ATSCV method with a case study on overtaking scenarios. Finally, Section V concludes the paper.

\section{Problem Formulation}

\subsection{Overtaking Scenarios}

\begin{figure}[!t]
  \centering
  \includegraphics[width=8.85cm]{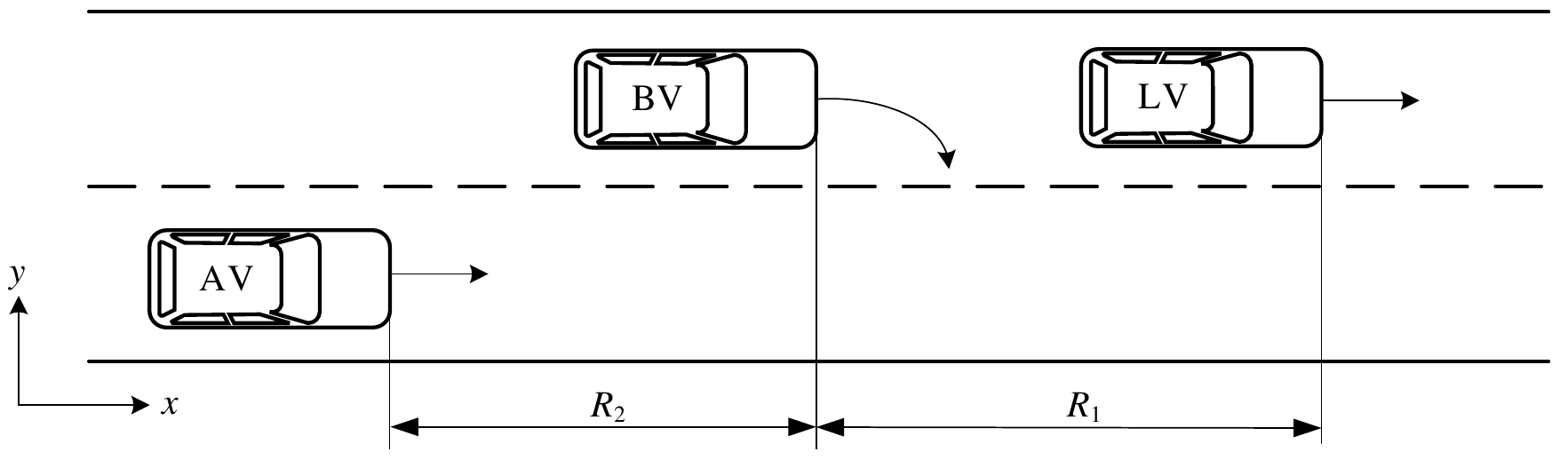}
  \caption{Illustration of the overtaking scenarios.}
  \label{fig:overtaking}
\end{figure}

The overtaking scenarios are shown in Fig.~\ref{fig:overtaking}. The key to handle the TSLG problem is to model the overtaking scenarios as Markov decision processes (MDPs). In overtaking scenarios, the leading vehicle (LV) in the left lane runs at a constant speed, whose state (including longitudinal position, lateral position and longitudinal velocity) and action (i.e., acceleration) are denoted as $s_{\mathrm{LV}}=( x_{\mathrm{LV}}, y_{\mathrm{LV}}, v_{\mathrm{LV}})$ and $a_{\mathrm{LV}}=0$, respectively. The background vehicle (BV) follows LV, whose state is denoted as $s_{\mathrm{BV}}=\left( x_{\mathrm{BV}}, y_{\mathrm{BV}}, v_{\mathrm{BV}} \right) $. Let $\mathcal{A}$ be the action space (i.e., acceleration space), and $\mathcal{A}^+\triangleq \mathcal{A} \cup \{\Rsh\} $ be the total action space, where $\Rsh$ represents right lane change. The action of BV $a_{\mathrm{BV}}\in \mathcal{A}^+$. Specifically, $a_{\mathrm{BV}}\in \mathcal{A}$ before BV cuts in, $a_{\mathrm{BV}}={\Rsh}$ when BV cuts in and after that $a_{\mathrm{BV}}=0$. The automated vehicle (AV) in the right lane runs at a constant speed before BV cuts in and follows BV after that, whose state is denoted as $s_{\mathrm{AV}}=\left( x_{\mathrm{AV}}, y_{\mathrm{AV}}, v_{\mathrm{AV}} \right) $. The action of AV $a_{\mathrm{AV}}=0$ before BV cuts in, and $a_{\mathrm{AV}}\in\mathcal{A}$ after that.

The state of an overtaking scenario is the collection of three vehicles' states $(s_{\mathrm{LV}},s_{\mathrm{BV}},s_{\mathrm{AV}})$. Since these state variables have interdependencies, we can formulate the state as 
\begin{equation}
  s\triangleq\left( v_{\mathrm{BV}}, R_1, \dot{R}_1, R_2, \dot{R}_2 \right) \in \mathcal{S},
\end{equation}
where $\mathcal{S} $ is the set of all feasible states, 
\begin{equation}
  \begin{aligned}
    R_1&=x_{\mathrm{LV}}-x_{\mathrm{BV}},\quad\dot{R}_1=v_{\mathrm{LV}}-v_{\mathrm{BV}},\\
    R_2&=x_{\mathrm{BV}}-x_{\mathrm{AV}},\quad\dot{R}_2=v_{\mathrm{BV}}-v_{\mathrm{AV}}.
  \end{aligned}
\end{equation}
The action of the overtaking scenario is defined as the action of BV, that is, $a\triangleq a_{\mathrm{BV}}\in\mathcal{A}^+$. Then an overtaking scenario is given by
\begin{equation}
  x=\left( s_0, a_0, s_1, a_1, \dots , s_m, a_m \right) \in \mathcal{X},
\end{equation}
where $\mathcal{X} $ is the set of all feasible scenarios, 
\begin{equation}
  s_k=\left( v_{\mathrm{BV},k}, R_{1,k}, \dot{R}_{1,k}, R_{2,k}, \dot{R}_{2,k} \right),~k=0,1,\dots,m
\end{equation}
is the state and $a_k=a_{\mathrm{BV}, k}$ is the action of BV at $k^{\mathrm{th}}$ time step, $0\leqslant m\leqslant M$, where $M$ is the maximum time step. When AV rear-ends BV after BV cuts in or BV does not cut in until AV passes BV, the overtaking scenarios terminate.

Suppose that the overtaking scenarios are simulated for 10 seconds at a frequency of 10 Hz, then the dimension of the scenario will exceed 10\textsuperscript{3}, resulting in the CoD problem. We note that the overtaking scenarios are significantly different from the simple scenarios that are usually studied such as car-following scenarios \cite{feng2021partii,zhao2017accelerated} and cut-in scenarios \cite{feng2020safety,feng2021partii,zhao2016accelerated}, where the accident events rely heavily on the initial states. On the contrary, the accident event of overtaking scenarios is more stochastic and complicated, because the right lane-changing action of BV is probabilistic and BV may have many chances of time steps to cut in, which essentially results in different cut-in and car-following scenarios. That is also the reason why the cut-in scenarios are usually low dimensional, while the overtaking scenarios are much more high dimensional. Moreover, the riskiest scenarios for the upcoming Level 3 automated lane keeping system (ALKS) will arguably be the overtaking scenarios \cite{unece2021uniform}. Therefore, the overtaking scenarios are selected for the case study.

\subsection{Testing Scenario Library Generation}

Let $\Omega =\mathcal{X}$ be the sample space, including all feasible overtaking scenarios. Consider the probability space $(\Omega, \mathcal{F}, \mathbb{P})$, where $\mathcal{F}$ is a $\sigma$-algerbra of subsets of $\Omega$ and $\mathbb{P}$ is a probability measure on $\mathcal{F}$. Since the scenarios are discretized, we choose $\mathcal{F} =\mathcal{P}(\Omega)$ to be the power set of $\Omega$. Then an event $A\subset\Omega$ is a subset of scenarios in the sample space $\Omega$. Let $X:x\mapsto x$, $x\in\mathcal{X}$ be the random variable of scenarios. To verify and validate AV, the accident event is usually the event of interest, which is defined as $A=\left\{ x\in \mathcal{X} \,\,| R_{1,m}\leqslant d_{\mathrm{accid}} \right\} $, where $d_{\mathrm{accid}}$ is the distance threshold for accident. The accident rate $\mu =\mathbb{P}(A)$ is utilized as the performance index of AV, which can also be represented as an expectation, i.e.,
\begin{equation}
  \mu
  =\mathbb{E}_p[\mathbb{I}_A(X)]
  =\sum_{x\in\mathcal{X}}\mathbb{P}(A|x)p(x),
\end{equation}
where $\mathbb{I}_A(X)$ is the indicator function of the event $A$,
\begin{equation}
  \mathbb{I}_A(X) =
  \begin{cases}	
    1, &\mathrm{if}~X\in A,\\	
    0, &\mathrm{if}~X\notin A,\\
  \end{cases}
\end{equation}
and $p$ is the naturalistic joint distribution of $x$. Assuming the Markovian property, the joint distribution can be factorized as
\begin{equation}
  p(x)=p(s_0)\prod_{k=0}^m p(a_k|s_k).
\end{equation}

The essence of testing AV in naturalistic driving environment (NDE) is to estimate the performance index $\mu$ by Monte Carlo simulation, i.e.,
\begin{equation}
  \hat{\mu}_n = \frac{1}{n}\sum_{i=1}^n\mathbb{P}(A|X_i), \quad X_i\sim p.
\end{equation}

The NDE suffers from the inefficiency problem, as the accident event is usually a rare event. To improve the efficiency of Monte Carlo simulation, importance sampling (IS) is adopted  to test AV with scenarios generated from a different probability distribution $q$, which is called the importance function and needs to satisfy $q(x)>0$ whenever $\mathbb{P}(A|x)p(x)>0$ \cite{zhao2016accelerated,feng2021parti,feng2021partii}. Since
\begin{equation}
  \begin{aligned}
    \mu
    &=\sum_{x\in\mathcal{X}}\mathbb{P}(A|x)p(x)
    =\sum_{x\in\mathcal{X}}\frac{\mathbb{P}(A|x)p(x)}{q(x)}q(x)\\
    &=\mathbb{E}_q\left[\frac{\mathbb{P}(A|X)p(X)}{q(X)}\right],\\
  \end{aligned}
\end{equation} 
the performance index can be estimated as
\begin{equation}
  \begin{aligned}
    \hat{\mu}_q
    &=\frac{1}{n}\sum_{i=1}^n\frac{\mathbb{P}(A|X_i)p(X_i)}{q(X_i)}, \quad X_i\sim q\\
    &=\frac{1}{n}\sum_{i=1}^n\mathbb{P}(A|X_i)\frac{p(s_{0,i})}{q(s_{0,i})}\prod_{k=0}^{m_i}\frac{p(a_{k,i}|s_{k,i})}{q(a_{k,i}|s_{k,i})},\\
  \end{aligned}
\end{equation}
where $X_i=(s_{0,i}, a_{0,i},\dots, s_{m_i,i}, a_{m_i,i})$, $i=1,\dots,n$.

\subsection{Naturalistic and Adversarial Driving Environment Generation}

The IS method is able to overcome the inefficiency problem with properly selected importance function and in fact, there exists optimal importance function whose estimation variance is zero \cite{feng2021parti}. However, the IS method faces the CoD problem if the testing scenarios are high-dimensional. To address both the inefficiency problem and the CoD problem, the naturalistic and adversarial driving environment (NADE) has been proposed to only sample critical variables with IS, while other variables remain its naturalistic distribution \cite{feng2021intelligent}.

Denote $x=(x_c, x_{-c})$, where $x_c=\{x_{c_1},\dots,x_{c_l}\}$, $l=0,1,\dots,L$ is the set of critical variables, $c_1,\dots,c_l$ are called the critical moments, $L$ is the maximum number of control steps among all scenarios $x\in\mathcal{X}$ and $x_{-c}$ is the set of other variables. Let $X_c:x\mapsto x_c$ be the random variable of critical variables and $X_{-c}:x\mapsto x_{-c}$ be the random variable of other variables. Then we have $X=(X_c, X_{-c})$. The random variable $X_c$ controls how to identify critical variables, and thus it's also called the control policy. The importance function can then be formulated as $q(x)=q(x_c)p(x_{-c})$, and therefore the performance index can be estimated in NADE as
\begin{equation}
  \begin{aligned}
    \tilde{\mu}_q
    &=\frac{1}{n}\sum_{i=1}^n\frac{\mathbb{P}(A|X_i)p(X_{c,i})}{q(X_{c,i})}, \quad X_i\sim q\\
    &=\frac{1}{n}\sum_{i=1}^n\mathbb{P}(A|X_i)\frac{p(s_{0,i})}{q(s_{0,i})}\prod_{k\in\mathcal{T}_{c,i}}\frac{p(a_{k,i}|s_{k,i})}{q(a_{k,i}|s_{k,i})},\\
  \end{aligned}
\end{equation}
where $X_{c,i}$ is the random variable of critical variables of $X_i$, and $\mathcal{T}_{c,i}$ is the set of critical moments of the $i^{\mathrm{th}}$ test.

It can be shown that with appropriate control policies and importance functions, NADE is able to greatly address the CoD problem and increase the evaluation efficiency \cite{feng2021intelligent}.

\section{Adaptive Testing with Sparse Control Variates}

\subsection{Control Variates}

Control variates (CVs) can be usefully combined with the mixture importance sampling. In mixture IS, we sample $X_1,\dots,X_n$ from the mixture importance function $q_\alpha=\sum_{j=1}^J\alpha_jq_j$, where $\alpha_j\geqslant0$, $\sum_{j=1}^J\alpha_j=1$ and the $q_j$ are importance functions. A control variate is a vector
\begin{equation}
  h(x)=(h_1(x),\dots,h_J(x))^\top
\end{equation}
for which $\sum_{x\in\mathcal{X}}h(x)=\theta$, where $\theta$ is a known value. Using the importance functions as control variates provides at least as good a variance reduction as we get from ordinary importance sampling \cite{owen2013monte}. When we combine mixture IS from $q_\alpha$ with control variates based on the component densities, the estimated performance index is
\begin{equation}
  \label{eq:MIS_CV}
  \hat{\mu}_{q_\alpha,\beta}=\frac{1}{n}\sum_{i=1}^n\frac{\mathbb{P}(A|X_i)p(X_i)-\sum_{j=1}^J\beta_jq_j(X_i)}{\sum_{j=1}^J\alpha_jq_j(X_i)} + \sum_{j=1}^J\beta_j
\end{equation}
for $X_i\sim q_\alpha$, where $\beta_j$ are control parameters.

We can compare the variance of mixture importance sampling to that of importance sampling with the individual mixture components $q_j$. Let $\beta^*$ be any minimizer over $\beta$ of $\mathrm{Var}_{q_\alpha}(\hat{\mu}_{q_\alpha,\beta})$. It can be proved \cite{owen2000safe} that
\begin{equation}
  \mathrm{Var}_{q_\alpha}(\hat{\mu}_{q_\alpha,\beta^*})\leqslant
  \min_{1\leqslant j\leqslant J} \frac{\sigma_{q_j}^2}{n\alpha_j},
\end{equation}
where
\begin{equation}
  \sigma_{q_j}^2=
  \mathrm{Var}_{q_j}\left(\frac{\mathbb{P}(A|X)p(X)}{q_j(X)}\right),~j=1,\dots,J.
\end{equation}
If any one of the $q_j$ is optimal then we will get a zero variance estimator of $\mu$. In practice, the optimal value $\beta^*$ is not known, and we can estimate it by multiple linear regression (MLR). Letting $Y_i=\mathbb{P}(A|X_i)p(X_i)/q_\alpha(X_i)$, $i=1,\dots,n$ and $Z_{ij}=q_j(X_i)/q_\alpha(X_i)-1$, $i=1,\dots,n$, $j=1,\dots,J-1$, then the $\beta^*$ can be estimated as the coefficients obtained from MLR of $Y_i$ on $Z_{ij}$.

\subsection{CoD of Control Variates}

Considering the Markov chain structure of overtaking scenarios, the mixture importance function of states is
\begin{equation}
  q_{\alpha}(s)
  =\sum_{j=1}^J {\alpha_jq_j(s)},~\forall s\in\mathcal{S},
\end{equation}
and the mixture importance function of actions given certain state $s\in\mathcal{S}$ is
\begin{equation}
  q_{\alpha}(a|s)
  =\sum_{j=1}^J{\alpha_jq_j(a|s)},~\forall a\in\mathcal{A}^+.
\end{equation}
Therefore, the mixture importance function of the overtaking scenarios is
\begin{equation}
  q_\alpha(x)=q_\alpha(s_0)\prod_{k=0}^m q_\alpha(a_k|s_k),~\forall x\in\mathcal{X},
\end{equation}
which is the product of $m+2$ individual mixture importance functions. According to Eq.~(\ref{eq:MIS_CV}), the control variates are
\begin{equation}
  q_{j_0,\dots,j_{m+1}}(x)=
  q_{j_0}(s_0)q_{j_1}(a_0|s_0)\cdots q_{j_{m+1}}(a_m|s_m),
\end{equation}
where $j_0,\dots,j_{m+1}=1,\dots,J$. Denote the corresponding control parameters as $\beta_{j_0,\dots,j_{m+1}}$.

The number of control parameters is $J^{m+2}$, which will increase exponentially with the dimension of scenarios, leading to the CoD problem when solving optimal control parameters via least squares method. 

\subsection{Sparse Control Variates}

To address the CoD problem of CVs, we propose the adaptive testing with sparse control variates (ATSCV) method to apply sparse control variates (SCVs) for testing results in NADE. The SCVs are constructed by the importance functions of critical variables of testing scenarios. We use ``sparse'' to describe these CVs because the critical variables are sparse in NADE. The number of critical variables can be limited to a much smaller number than the whole dimension of scenarios, and so can the number of SCVs, which will greatly address the CoD problem of CVs.

Using scenarios sampled from $q_\alpha$, the performance index can be estimated in NADE as
\begin{equation}
  \label{eq:MIS_NADE}
  \tilde{\mu}_{q_\alpha}
  =\frac{1}{n}\sum_{i=1}^n\frac{\mathbb{P}(A|X_i)p(X_{c,i})}{q_\alpha(X_{c,i})}, \quad X_i\sim q_\alpha.
\end{equation}
Let $\mathcal{X}_l=\{x\in\mathcal{X}:|x_c|=l\}$, $l=0,1,\dots,L$ be the set of scenarios that are controlled $l$ steps, satisfying $\bigcup_{l=0}^L\mathcal{X}_l=\mathcal{X}$. Denote the performance index of scenarios over $\mathcal{X}_l$ as
\begin{equation}
  \mu_l\triangleq\mathbb{E}_p[\mathbb{I}_A(X)\mathbb{I}_{\mathcal{X}_l}(X)],~l=0,1,\dots,L.
\end{equation}
Similar to Eq.~(\ref{eq:MIS_NADE}), the estimation of $\mu_l$ can be written as
\begin{equation}
  \tilde{\mu}_{l,q_\alpha}
  =\frac{1}{n}\sum_{i=1}^n\frac{\mathbb{P}(A|X_i)\mathbb{I}_{\mathcal{X}_l}(X_i)p(X_{c,i})}{q_\alpha(X_{c,i})},
\end{equation}
then we have
\begin{equation}
  \mu=\sum_{l=0}^L\mu_l,
\end{equation}
and
\begin{equation}
  \begin{aligned}
    \tilde{\mu}_{q_\alpha}
    &=\sum_{l=0}^L\frac{1}{n}\sum_{i=1}^n\frac{\mathbb{P}(A|X_i)\mathbb{I}_{\mathcal{X}_l}(X_i)p(X_{c,i})}{q_\alpha(X_{c,i})}
    =\sum_{l=0}^L\tilde{\mu}_{l,q_\alpha}.
  \end{aligned}
\end{equation}

Let $q_{j_1,\dots,j_l}(x_c)=q_{j_1}(x_{c_1})\cdots q_{j_l}(x_{c_l})$ be the importance functions of critical variables that sample $x_{c_1},\dots,x_{c_l}$ from $q_{j_1},\dots,q_{j_l}$ respectively, where $j_1,\dots,j_l=1,\dots,J$, $l=1,\dots,L$. We deal with the CoD problem by applying $q_{j_1,\dots,j_l}$ instead of $q_{j_0,\dots,j_{m+1}}$ as SCVs to the estimation $\tilde{\mu}_{l,q_\alpha}$. If the control steps $l\ll m$, then the number of control variates can be reduced considerably, which will greatly address the CoD problem. For a good enough control policy, this assumption can be guaranteed. Since the numbers of SCVs with different control steps are different, we group the testing results by their control steps and apply the MLR within each group.

Denote the summation of the product of control variates and corresponding control parameters as
\begin{equation}
  h_l(x_c)\triangleq
  \sum_{j_1,\dots,j_l}\beta_{j_1,\dots,j_l}q_{j_1,\dots,j_l}(x_c),
  ~l=1,\dots,L,
\end{equation}
and the summation of $h_l(x_c)$ over $x\in\mathcal{X}_l$ as
\begin{equation}
  \theta_l\triangleq\sum_{x\in\mathcal{X}_l}h_l(x_c),
  ~l=1,\dots,L.
\end{equation}
With the SCVs $q_{j_1,\dots,j_l}$, the estimation $\tilde{\mu}_{l,q_\alpha}$ can be adaptively evaluated as
\begin{equation}
  \label{eq:ATSCV}
  \tilde{\mu}_{l,q_\alpha,\beta_l}
  =\frac{1}{n}\sum_{i=1}^n\frac{\mathbb{P}(A|X_i)p(X_{c,i})-h_l(X_{c,i})}{q_\alpha(X_{c,i})}\mathbb{I}_{\mathcal{X}_l}(X_i)+\theta_l
\end{equation}
for $l=1,\dots,L$. Note that for $l=0$, the scenarios are sampled from naturalistic distribution $p$, and therefore we do not apply SCVs to its estimation, i.e.,
\begin{equation}
  \tilde{\mu}_{0,q_\alpha,\beta_0}=\frac{1}{n}\sum_{i=1}^n\mathbb{P}(A|X_i)\mathbb{I}_{\mathcal{X}_0}(X_i).
\end{equation}
Therefore, the performance index estimated by the ATSCV method is
\begin{equation}
  \tilde{\mu}_{q_\alpha,\beta}=
  \sum_{l=0}^L\tilde{\mu}_{l,q_\alpha,\beta_l}.
\end{equation}

\section{Simulation Analysis}

\subsection{Generation of NDE}

The goal of NDE is to generate human-like driving behaviors with probability distributions consistent with the naturalistic driving data (NDD). Since the overtaking scenarios are modeled with MDPs, after generating the initial state, the actions of vehicles are determined by current states, which can be sampled from empirical distributions obtained from NDD. The decision process can be represented by a decision tree, where each node denotes vehicle states and each path denotes a specific realization of vehicle behaviors. For simplicity, as a replacement of empirical distributions, we use intelligent driver model (IDM) \cite{ro2017formal} to model car-following scenarios and stochastic minimizing overall braking induced by lane changes (MOBIL) model \cite{feng2021intelligent} to govern lane-changing behaviors. All vehicles select actions independently and simultaneously for each time step (0.1 s).

The initial state of overtaking scenarios is set as
\begin{equation}
  s_0=[8, R_1, -5, 5, -5],~R_1\sim\mathcal{U}(30,32),
\end{equation}
where $\mathcal{U}$ is the uniform distribution. The overtaking scenario has two stages separated by the time when BV cuts in. Before cutting in, BV is controlled by IDM and stochastic MOBIL model, while LV and AV run at a constant speed. After cutting in, AV is controlled by IDM, while LV and BV run at a constant speed. The cut-in maneuver is set completed within one time step. The simulation continues until AV rear-ends BV or AV passes BV.

\subsection{Generation of NADE}

The key of NADE generation is to construct new distributions as the replacement of the naturalistic distributions in NDE, for sampling actions of BV \cite{feng2021intelligent}. In NADE, we only twist the action distribution of BV at critical moments, while keeping naturalistic distribution as in NDE at other time steps. To construct the importance function, at each time step, the action of BV will be evaluated by maneuver criticality, which is computed as a multiplication of exposure frequency $p(a|s)$ and maneuver challenge $\mathbb{P}(A|s,a)$, i.e.,
\begin{equation}
  V(a|s)=\mathbb{P}(A|s,a)p(a|s).
\end{equation}
The exposure frequency represents the probability of the action $a$ given the state $s$ in NDE, while the maneuver challenge measures the accident probability of AV given the state-action pair $(s,a)$.

Let $p_R$ be the probability of right lane change induced from stochastic MOBIL model, then the exposure frequency is
\begin{equation}
  p(a|s)=
  \begin{cases}
    p_R,&\mathrm{if}~a={\Rsh},\\
    1-p_R,&\mathrm{if}~a=\mathrm{IDM}(s)\in\mathcal{A},\\
    0,&\mathrm{otherwise},\\
  \end{cases}
\end{equation}
where $\mathrm{IDM}(s)$ is the acceleration calculated by IDM given the state $s$. Since the AV model is usually a black-box, we use surrogate models (SMs) to approximate the maneuver challenge. In this paper, IDM and full velocity difference model (FVDM) \cite{ro2017formal} are selected as SMs with adjusted parameters as follows: 1) IDM. 2) FVDM-I. 3) FVDM-II with $\kappa=6$.

Let $S_j$, $j=1,\dots,J$ denotes the accident event between BV and the $j^{\mathrm{th}}$ SM, where $J=3$, then the maneuver challenge can be approximated by $\mathbb{P}(S_j|s,a)$, $j=1,\dots,J$. If BV follows LV with $a\in\mathcal{A}$, then we have
\begin{equation}
  \label{eq:maneuver_challenge}
  \mathbb{P}(S_j|s,a)
  =\sum_{a\in\mathcal{A}^+}{\mathbb{P}(S_j|s',a) p(a|s')},
\end{equation}
where $s'$ is the next state of the overtaking scenario given the state-action pair $(s,a)$. The maneuver challenge $\mathbb{P}(S_j|s,a)$ can be solved by recursion with Eq.~(\ref{eq:maneuver_challenge}) and the following two terminal conditions.
\begin{enumerate}
  \item If BV gets past by AV, then there will be no accident. Therefore, $\mathbb{P}(S_j|s,a)=0$ for $R_2<0$.
  \item After BV cuts in, whether the accident will occur is predicted by the $j^{\mathrm{th}}$ SM and the state after cutting in. Thus, for $a={\Rsh}$, $\mathbb{P}(S_j|s,a)=1$ if AV rear-ends BV, and otherwise $\mathbb{P}(S_j|s,a)=0$.
\end{enumerate}

\begin{algorithm}[!t]
  \label{alg:ATSCV}
  \caption{Adaptive testing with sparse control variates by multiple linear regression}
  \KwIn{$p$, $q_\alpha$, $X_{c,i}$, and $\mathbb{P}(A|X_i)$, $i=1,\dots,n$}
  \KwOut{$\tilde{\mu}_{q_\alpha,\hat{\beta}}$}
  initialize $Y_l$ and $Z_l$ as empty arrays, $l=0,\dots,L$\;
  \For{$i\leftarrow 1$ \KwTo $n$}{
    $l\leftarrow$ number of control steps of $X_{c,i}$\;
    \eIf{$l=0$}{
      append $Y_l$ with $\mathbb{P}(A|X_i)$\;
      append $Z_{l}$ with 0\;
    }{
      append $Y_l$ with $\mathbb{P}(A|X_i)p(X_{c,i})/q_\alpha(X_{c,i})$\;
      append $Z_{l}$ with $\mathrm{vec}(q_{j_1,\dots,j_l}(X_{c,i})/q_\alpha(X_{c,i}))$, $j_1,\dots,j_l=1,\dots,J-1$\; 
      /* $\mathrm{vec}(\cdot)$ is the vectorization operator */
    }
  }
  \For{$l\leftarrow 0$ \KwTo $L$}{
    $Z_l\leftarrow Z_l-\mathrm{average}(Z_l)$\;
    $\mathrm{MLR}\leftarrow$ multiple linear regression of $Y_l$ on $Z_{l}$\;
    $\hat{\beta}_l\leftarrow$ estimated coefficients from $\mathrm{MLR}$\;
    $\hat{\eta}_l\leftarrow$ estimated intercept from $\mathrm{MLR}$\;
    $\tilde{\mu}_{l,q_\alpha,\hat{\beta}_l}\leftarrow\mathrm{length}(Y_l)\hat{\eta}_l/n$\;
  }
  return $\tilde{\mu}_{q_\alpha,\hat{\beta}}\leftarrow\sum_{l=0}^L\tilde{\mu}_{l,q_\alpha,\hat{\beta}_l}$\;
\end{algorithm}

The criticality of BV can then be calculated as the summation of maneuver criticality over all actions, i.e.,
\begin{equation}
  C_j(s)=\sum_{a\in\mathcal{A}^+}V_j(a|s),~j=1,\dots,J,
\end{equation}
where $V_j(a|s)=\mathbb{P}(S_j|s,a)p(a|s)$. If any $C_j(s)>0$, then the moment of state $s$ is defined as the critical moment and the action of BV will be adjusted by sampling from the mixture importance function $q_\alpha(a|s)=\sum_{j=1}^J\alpha_jq_j(a|s)$, where $\alpha_j=1/3$, $j=1,\dots,J$ is the weight of the $j^{\mathrm{th}}$ SM,
\begin{equation}
  q_j(a|s)=\begin{cases}
    \epsilon p(a|s)+(1-\epsilon)\frac{V_j(a|s)}{C_j(s)}, & \mathrm{if}~C_j(s)>0,\\
    p(a|s), &\mathrm{otherwise},
  \end{cases}
\end{equation}
where $\epsilon=0.1$ is the weight of the naturalistic distribution.

The ATSCV method can be applied while evaluating the testing results in NADE. Algorithm \ref{alg:ATSCV} describes this process. The key is to identify the critical variables of testing scenarios to formulate the sparse control variates for each testing result. Then the performance index of different control steps can be estimated as the intercept obtained from multiple linear regression of weighted testing results on corresponding sparse control variates.

\subsection{Evaluation Results}

\begin{figure}[!t]
  \centering
  \includegraphics[width=8.85cm]{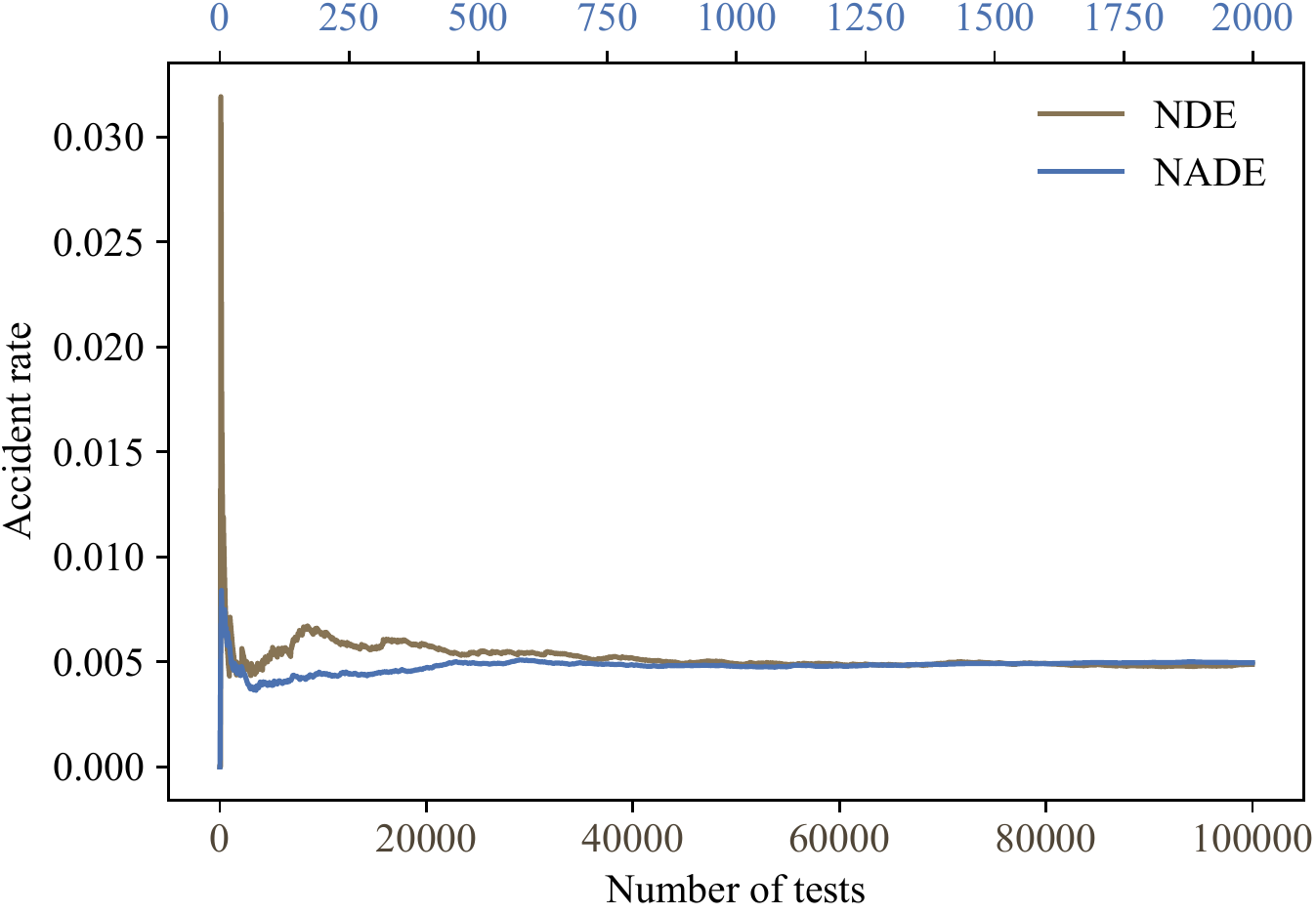}
  \caption{Accident rate of AV in NDE and NADE.}
  \label{fig:accident_rate_NDE_NADE}
\end{figure}

The accuracy and efficiency of AV evaluation in NDE and NADE are validated in our simulation. Fig.~\ref{fig:accident_rate_NDE_NADE} shows the evaluation results of the accident rate per test in NDE and NADE. The gray line represents the testing results in NDE, and the bottom $x$-axis indicates the number of tests. The blue line represents the testing results in NADE, and the top $x$-axis for the number of tests. It can be seen that NADE obtains the same accident rate estimation with NDE by a much smaller number of tests.

\begin{figure}[!t]
  \centering
  \includegraphics[width=8.85cm]{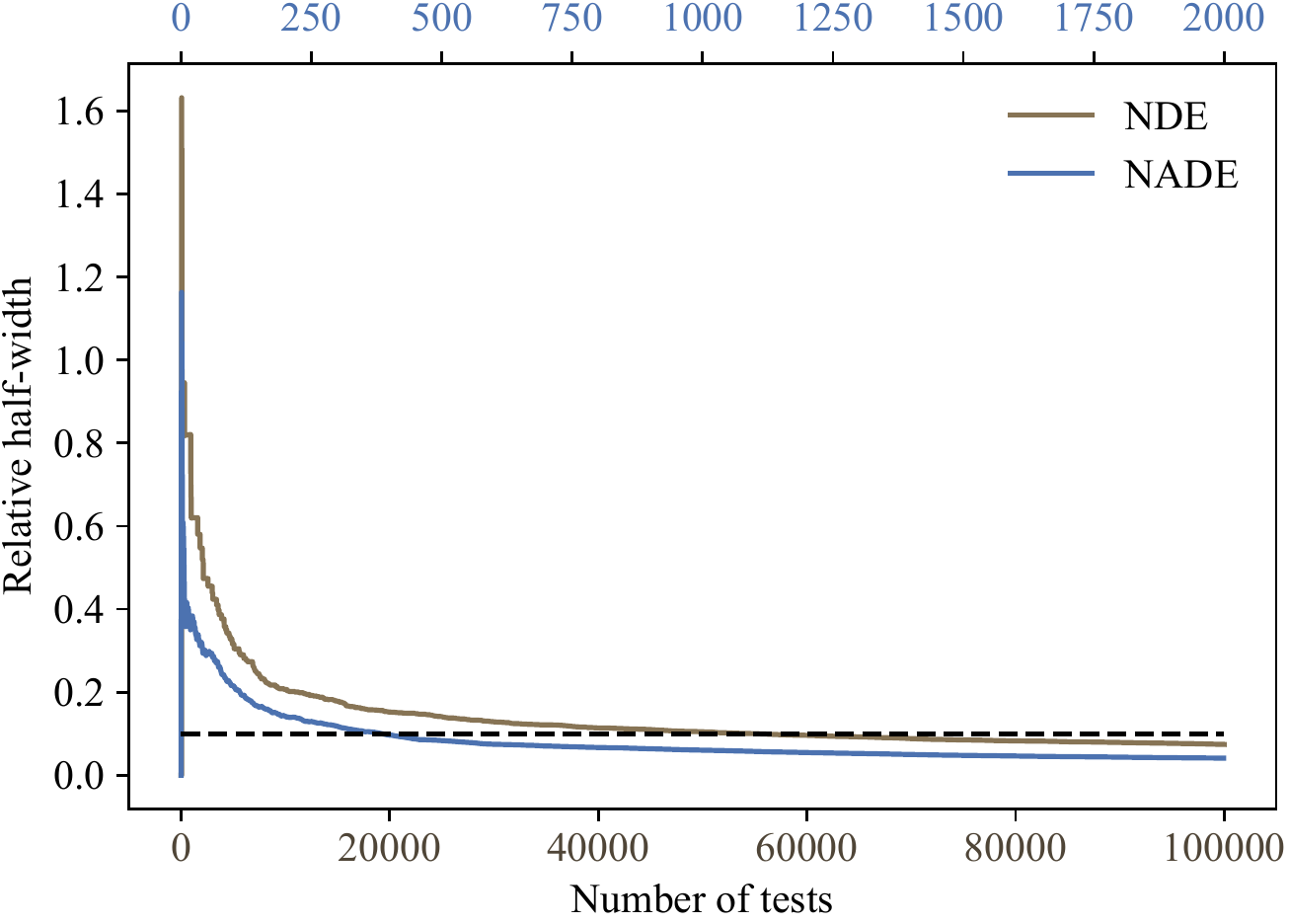}
  \caption{RHW of AV evaluation in NDE and NADE.}
  \label{fig:relative_half_width_NDE_NADE}
\end{figure}

The relative half-width (RHW) \cite{zhao2016accelerated} is selected to measure the evaluation precision. With the confidence level $100(1-\gamma)\%$, the RHW of $\tilde{\mu}_{q_\alpha,\beta}$ is defined as
\begin{equation}
  l_r=z_\gamma\frac{\sqrt{\mathrm{Var}(\tilde{\mu}_{q_\alpha,\beta})}}{\tilde{\mu}_{q_\alpha,\beta}},
\end{equation}
where $z_\gamma=\Phi^{-1}(1-\gamma/2)$, $\gamma=0.1$, and $\Phi^{-1}$ is the inverse cumulative distribution function of the standard normal distribution $\mathcal{N}(0,1)$. Fig.~\ref{fig:relative_half_width_NDE_NADE} presents the minimum number of tests for reaching a predetermined precision threshold (RHW is 0.1) in NDE and NADE. It can be found that NADE requires only 385 number of tests, while NDE requires 5.5\,$\times$\,10\textsuperscript{4} number of tests, resulting in an acceleration of evaluation for about 143 times.

\begin{figure}[!t]
  \centering
  \includegraphics[width=8.85cm]{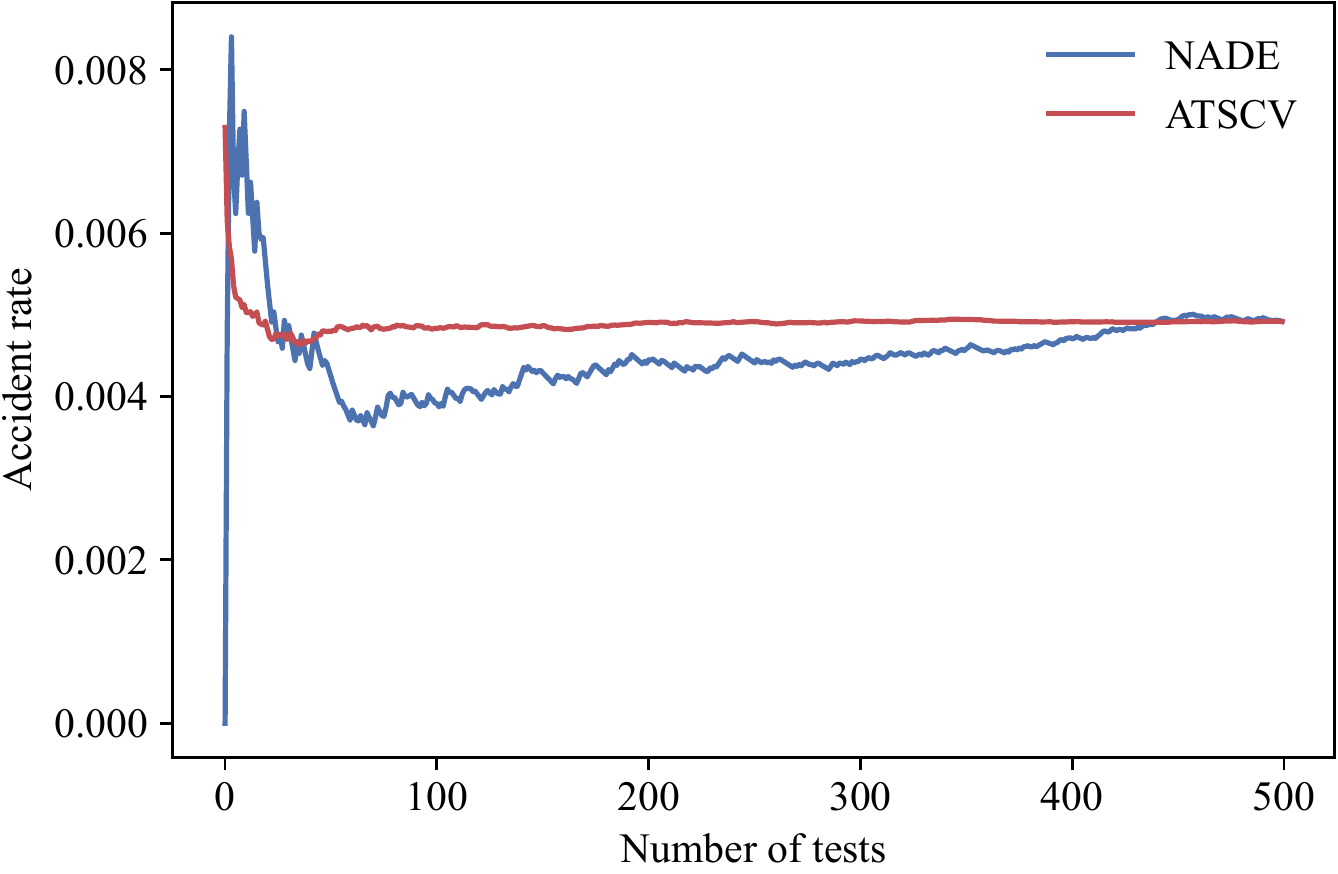}
  \caption{Accident rate of AV in NADE with and without SCVs.}
  \label{fig:accident_rate_NADE_CV}
\end{figure}

\begin{figure}[!t]
  \centering
  \includegraphics[width=8.85cm]{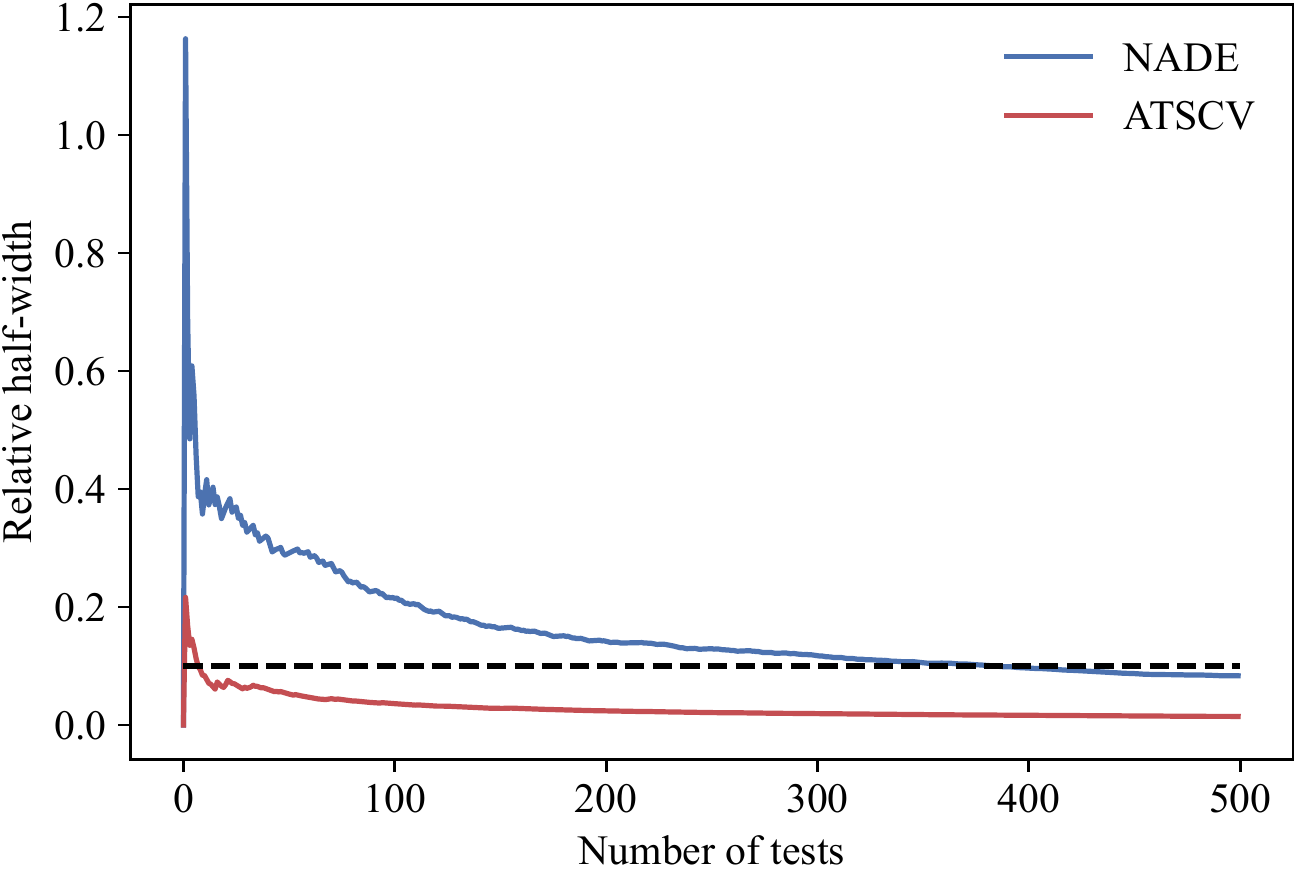}
  \caption{RHW of AV evaluation in NADE with and without SCVs.}
  \label{fig:relative_half_width_NADE_CV}
\end{figure}

\begin{figure}[!t]
  \centering
  \includegraphics[width=8.85cm]{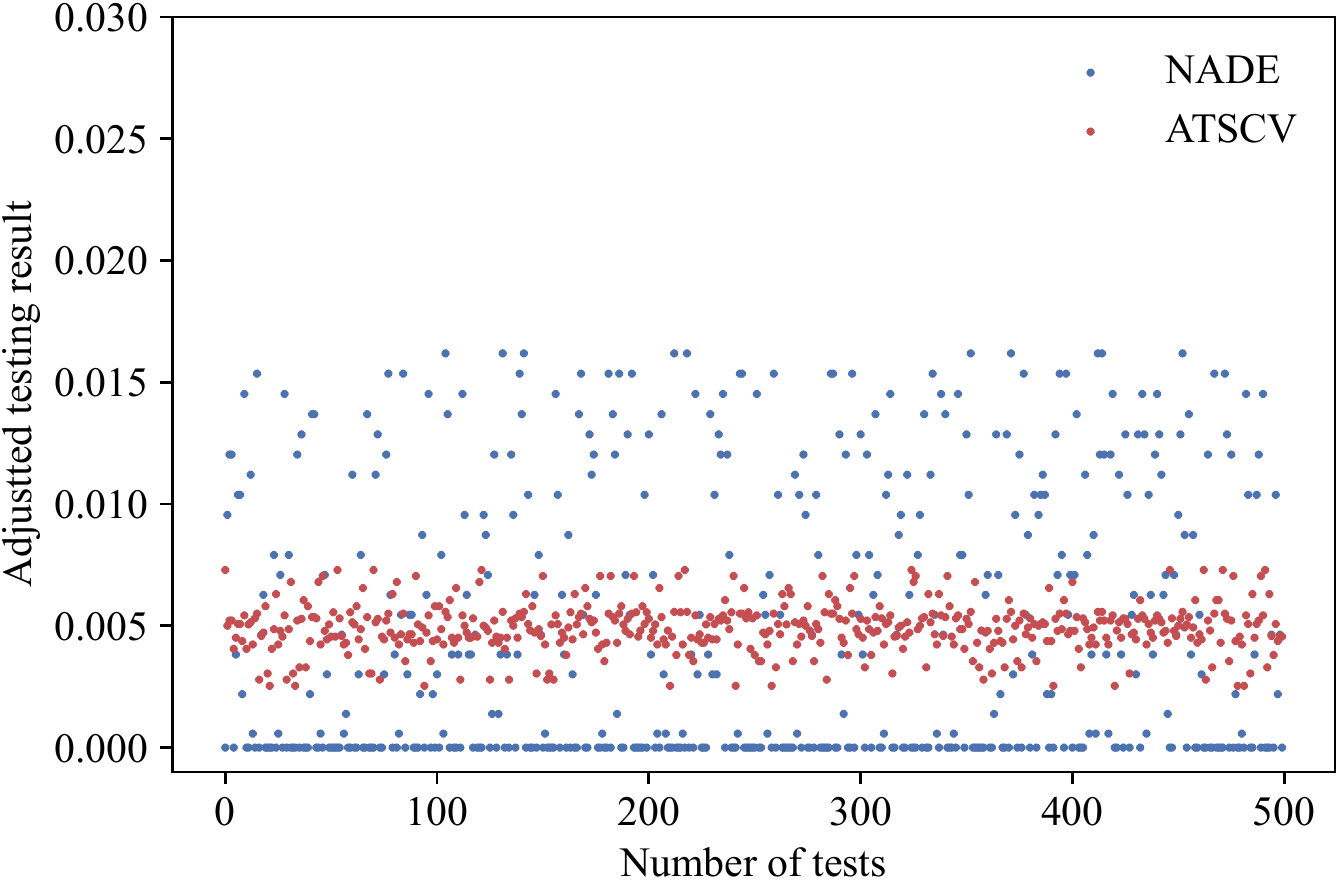}
  \caption{Adjusted testing results in NADE with and without SCVs.}
  \label{fig:accident_NADE_ATSCV}
\end{figure}

To investigate the influences of the ATSCV method, we compare the efficiency of AV evaluation in NADE with and without SCVs. As shown in Fig.~\ref{fig:accident_rate_NADE_CV}, NADE and ATSCV can converge to the same accident rate after a sufficient number of tests. To reach the 0.1 relative half-width, the total required number of tests is 385 and 12, respectively, as shown in Fig.~\ref{fig:relative_half_width_NADE_CV}. Therefore, the ATSCV method can further accelerate the evaluation by about 30 times comparing with the evaluation results in NADE. Fig.~\ref{fig:accident_NADE_ATSCV} presents the adjusted testing results in NADE with and without SCVs, where each blue point is the value
\begin{equation}
  \frac{\mathbb{P}(A|X_i)p(X_{c,i})}{q_\alpha(X_{c,i})},~i=1,\dots,n
\end{equation}
in Eq.~(\ref{eq:MIS_NADE}), and each red point represents the value
\begin{equation}
  \frac{\mathbb{P}(A|X_i)p(X_{c,i})-h_l(X_{c,i})}{q_\alpha(X_{c,i})}+\theta_l
\end{equation}
in Eq.~(\ref{eq:ATSCV}) for $X_i\in\mathcal{X}_l$, $l=1,\dots,L$ or $\mathbb{P}(A|X_i)$ for $X_i\in\mathcal{X}_0$. It can be seen that the ATSCV method is able to adjust the testing results into a much narrower interval, resulting in a considerable reduction of the estimation variance.

To further verify the reliability of the ATSCV method, the simulation is repeated 100 times. Results show that the average accelerated rate is 28.34 and the standard deviation is 6.27. The required numbers of tests in all 100 experiments are less than the evaluation results in NADE.

\section{Conclusion}

In this paper, the adaptive testing with sparse control variates (ATSCV) method is proposed to adaptively evaluate the testing results of connected and automated vehicles in complex scenarios. The major idea is to apply SCVs of critical variables for the evaluation of scenarios in NADE. With scenarios modeled as MDPs, the mixture importance functions are the summation of weighted defensive importance functions obtained using SMs. The SCVs are constructed as the importance functions of critical variables, each of which is the product of a combination of importance functions obtained by different SMs at critical moments. The overtaking scenarios are investigated for safety testing. Comparing with the evaluation results in NDE and NADE, the proposed ATSCV method is always more efficient. In the near future, the theoretical analysis with rigorous proofs of the ATSCV method will be developed and more realistic cases with large-scale naturalistic driving data will be studied.

\bibliographystyle{IEEEtran}
\bibliography{IEEEabrv,reference.bib}

\end{document}